Do disruption index indicators measure what they propose to measure?

The comparison of several indicator variants with assessments by peers

Lutz Bornmann*, Sitaram Devarakonda**, Alexander Tekles*+, George Chacko**

*Division for Science and Innovation Studies
Administrative Headquarters of the Max Planck Society
Hofgartenstr. 8,
80539 Munich, Germany.
Email: bornmann@gv.mpg.de
Email: alexander.tekles.extern@gv.mpg.de

**NET ESolutions Corporation
8100 Greensboro Dr
McLean, VA 22102, USA
Email: sitaram@nete.com
Email: netelabs@nete.com

+Ludwig-Maximilians-Universität Munich
Department of Sociology
Konradstr. 6
80801 Munich, Germany.

**Abstract**

Recently, Wu, Wang, and Evans (2019) and Bu, Waltman, and Huang (2019) proposed a new family of indicators, which measure whether a scientific publication is disruptive to a field or tradition of research. Such disruptive influences are characterized by citations to a focal paper, but not its cited references. In this study, we are interested in the question of convergent validity, i.e., whether these indicators of disruption are able to measure what they propose to measure ('disruptiveness'). We used external criteria of newness to examine convergent validity: in the post-publication peer review system of F1000Prime, experts assess papers whether the reported research fulfills these criteria (e.g., reports new findings). This study is based on 120,179 papers from F1000Prime published between 2000 and 2016. In the first part of the study we discuss the indicators. Based on the insights from the discussion, we propose alternate variants of disruption indicators. In the second part, we investigate the convergent validity of the indicators and the (possibly) improved variants. Although the results of a factor analysis show that the different variants measure similar dimensions, the results of regression analyses reveal that one variant ($DI_5$) performs slightly better than the others.

# 1 Introduction

Citation analyses often focus on counting the number of citations to a focal paper (FP). To assess the academic impact of the FP, its citation count is compared with the citation count for a similar paper (SP, which has been published in the same research field and year). If the FP receives significantly more citations than the SP, its impact is noteworthy: the FP seems to be more useful or interesting for other researchers than the SP. However, this simple counting and comparing of citations do not reveal what the reasons for the impact of publications might be. As the overviews by Bornmann and Daniel (2008) and Tahamtan and Bornmann (2019) show various reasons exist why publications are (highly) cited. Especially for research evaluation purposes, it is very interesting to know whether certain publications have impact because they report novel or revolutionary results. These are the results from which science (and society) mostly profit.

In this paper, we focus on a new type of indicator family measuring impact of publications by examining not only the number of citations received, but also the references cited in publications. Recently, Wu et al. (2019) and Bu et al. (2019) proposed a new family of indicators, which measure whether a FP disrupts a field or tradition of research. Azoulay (2019) describes the so called disruption index proposed by Wu et al. (2019) as follows: "when the papers that cite a given article also reference a substantial proportion of that article's references, then the article can be seen as consolidating its scientific domain. When the converse is true – that is, when future citations to the article do not also acknowledge the article's own intellectual forebears – the article can be seen as disrupting its domain" (p. 331).

We are interested in the question of whether disruption indicators are able to measure what they propose to measure. The current study has two parts: in the first part, we discuss the introduced indicators by Wu et al. (2019) and Bu et al. (2019) and identify possible weaknesses. Based on the insights from the discussion, we propose alternate variants of disruption indicators. In the second part, we investigate the convergent validity of the indicators proposed by Wu et al. (2019) and Bu et al. (2019) and the (possibly) improved variants. We used an external criterion of newness, which is available at the paper level for a large paper set: tags (e.g., "new finding") assigned to papers by peers expressing newness.

Convergent validity asks "to what extent does a bibliometric exercise exhibit externally convergent and discriminant qualities? In other words, does the indicator satisfy the condition that it is positively associated with the construct that it is supposed to be measuring? The criteria for convergent validity would not be satisfied in a bibliometric experiment that found little or no correlation between, say, peer review grades and citation measures" (Rowlands, 2018). The analyses are intended to identify the indicator (variant), which is strongly related to assessments by peers (concerning newness) than other indicators.

## 2 Indicators measuring disruption

The new family of indicators measuring disruption has been developed based on the previous introduction of another indicator family measuring novelty. Research on the novelty indicator family is based on the view of research as a "problem solving process involving various combinatorial aspects so that novelty comes from making unusual combinations of preexisting components" (Wang, Lee, & Walsh, 2018, p. 1074). Uzzi, Mukherjee, Stringer, and Jones (2013) analyzed cited references, and investigated whether referenced journal pairs in papers are atypical or not. Papers with many atypical journal pairs were denoted as papers with high novelty potential. The authors argue that highly-cited papers are not only highly novel, but are also very conventionally oriented. In a related study, Boyack and Klavans (2014) reported strong disciplinary and journal effects in inferring novelty.

In recent years, Lee, Walsh, and Wang (2015) proposed an adapted version of the novelty measure proposed by Uzzi et al. (2013); Wang, Veugelers, and Stephan (2017) and Stephan, Veugelers, and Wang (2017) introduced a novelty measure focusing on publications with great potential of being novel by identifying new journal pairs (instead of atypical pairs). Other studies in the area of measuring novelty have been published by Foster, Rzhetsky, and Evans (2015), Foster et al. (2015), Boudreau, Guinan, Lakhani, and Riedl (2016), Carayol, Lahatte, and Llopis (2017), Mairesse and Pezzoni (2018), Bradley et al. (2019), and Wagner, Whetsell, and Mukherjee (2019) each with different focus. According to the conclusion by Wang et al. (2018), "prior work suggests that coding for rare combinations of prior knowledge in the publication produces a useful a priori measure of the novelty of a scientific publication" (p. 1074).

Novelty indicators have been developed against the backdrop of the desire to identify and measure creativity. How is creativity defined? According to Hemlin, Allwood, Martin, and Mumford (2013) "creativity is held to involve the production of high-quality, original, and elegant solutions to complex, novel, ill-defined, or poorly structured problems" (p. 10). Puccio, Mance, and Zacko-Smith (2013) claim that "many of today's creativity scholars now define creativity as the ability to produce original ideas that serve some value or solve some problem" (p. 291). The connection between the indicators measuring novelty and creativity of research is made by that stream of research viewing creativity "as an evolutionary search process across a combinatorial space and sees creativity as the novel recombination of elements" (Lee et al., 2015, p. 685). For Estes and Ward (2002) "creative ideas are often the result of attempting to determine how two otherwise separate concepts may be understood together" (p. 149) whereby the concepts may refer to different research traditions or disciplines. Similar statements on the roots of creativity can be found in the literature from other authors as the overview by Wagner et al. (2019) shows. Bibliometric novelty indicators try to capture the combinatorial dynamic of papers (and thus, the creative potential of papers, see Tahamtan & Bornmann, 2018) by investigating lists of cited references for new or unexpected combinations (Wagner et al., 2019).

In a recent study, Bornmann, Tekles, Zhang, and Ye (in press) investigated two novelty indicators and tested whether they exhibit convergent validity. They used a similar design as this study and found that only one indicator is convergent valid.

In this context of measuring creativity, not only has the development of indicators measuring novelty occurred, but also the introduction of indicators identifying disruptive research. These indicators are interested in exceptional research which turns knowledge formation in a field around. The family of disruptive indicators proposed especially by Wu et al. (2019) and Bu et al. (2019) seizes on the concept of Kuhn (1962) who differentiated between phases of normal science and scientific revolutions. Normal science is characterized by paradigmatic thinking, which is rooted in traditions, and consensus orientation; scientific revolutions follow divergent thinking and openness (Foster et al., 2015). Whereas normal science means linear accumulation of research results in a field (Petrovich, 2018), scientific revolutions are dramatic changes with an overthrow of established thinking (Casadevall & Fang, 2016). Preconditions for scientific revolutions are creative knowledge claims which disrupt linear accumulation processes in field-specific research (Kuukkanen, 2007).

Bu et al. (2019) see the development of disruptive indicators in the context of a multi-dimensional perspective on citation impact. In contrast to simple citation counting under the umbrella of a one-dimensional perspective, the multi-dimensional perspective considers breadth and depth through cited references of a FP and the cited references of its citing papers (see also Marx & Bornmann, 2016). In contrast to the family of novelty indicators, which are based exclusively on cited references, disruption indicators combine cited references of citing papers with cited references data of FPs. The disruptiveness of a FP is measured based on the extent to which the cited references of the papers citing the FP also refer to the cited references of the FP. According to this idea, many citing papers not referring to the FP's cited references indicate disruptiveness. In this case, the FP is the basis for new work which does not depend on the context of the FP, i.e. the FP gives rise to new research.

Disruptiveness was first described by Wu et al. (2019) and presented as a weighted index $DI_1$ (see Figure 1) calculated for a FP by dividing the difference between the number of publications that cite the FP without citing any of its cited references ($N_i$) and the number of publications that cite both the FP and at least one of its cited references ($N_j^1$) by the sum of $N_i$, $N_j^1$, and $N_k$ (the number of publications that cite at least one of the FP's cited reference without citing the paper itself). Simply put, the ratio of $\frac{N_i - N_j}{N_i + N_j + N_k}$. High positive values of this indicator should be able to point to disruptive research; high negative values should reflect developmental research (i.e., new research which continues previous research lines).

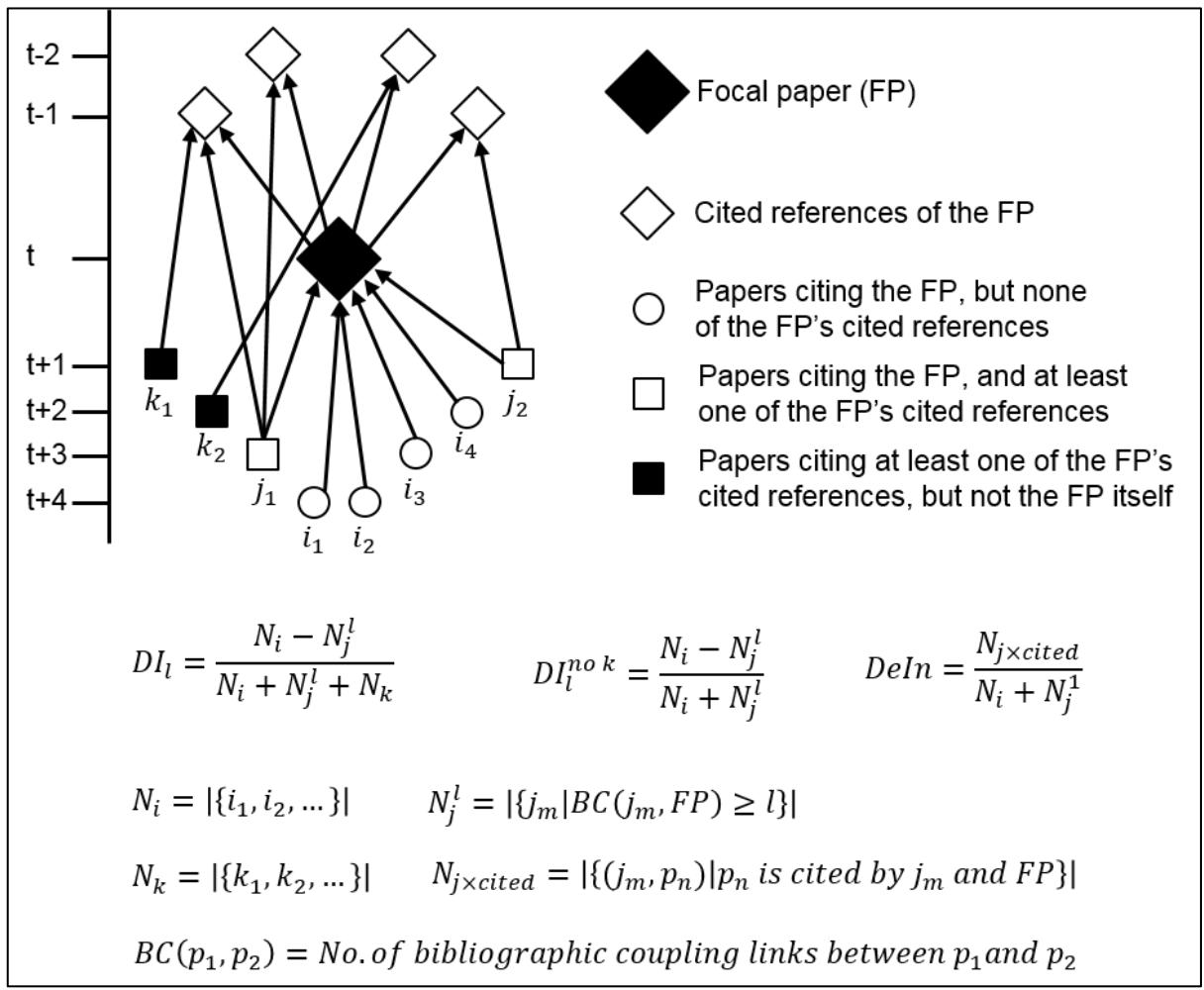

Figure 1. Different roles of papers in citation networks for calculating disruption indicators and formulae for different disruption indicators (see Wu & Wu, 2019)

Regarding $DI_1$, the function of the term $N_k$ is not entirely clear. Given that disruptiveness defined as the degree to which new papers are based on the FP but not on the papers that the FP is based on, its consideration in the formula may have unintended effects on the results (Wu & Wu, 2019). The impact of the focal paper's cited references may disproportionately affect the disruption score when $N_k$ is used since the focus is the degree to which new papers cite the FP, but not the papers that the FP is based on. For example, if a cited reference (paper) of the FP gets cited many times (by papers not citing the FP), then the magnitude of the disruption score may decrease considerably. This is true even if the citing papers of the FP do not change.

Another problem with the term $N_k$ is that it is often very large compared to the other terms in the formula (Wu & Yan, 2019). This produces very small disruption values in many cases, since $N_k$ only occurs in the denominator of the formula. As a result, the disruption index is very similar for many papers, and the measure can discriminate only for a few papers. Consequently, an alternative would be simply to drop the term $N_k$, which corresponds to $DI_1^{no\ k}$ according to the formula in Figure 1. This variant of the disruption indicator has been proposed by Wu and Yan (2019). With $\frac{N_i}{N_i+N_j}$, a very similar approach for calculating a paper's disruption has been proposed by Bu et al. (2019). This indicator can be defined as a function of $DI_1^{no\ k}$, such that differences between papers just change by the factor 0.5, so that both variants allow identical conclusions. In our analyses, we will consider $DI_1^{no\ k}$, because it has the same range of output values as the original disruption index $DI_1$.

In contrast to the aforementioned variants of indicators measuring disruptiveness, Bu et al. (2019) also proposed an indicator that considers the number of bibliographic coupling links between the FP and its citing papers' cited references, i.e., how many times the FP and one of its citing papers have a common cited reference. This approach takes into account how strong the FP's citing papers rely on the cited references of the FP, instead of just considering if this relationship exists (in the form of at least one citation link). The corresponding indicator proposed by Bu et al. (2019) (denoted as $DeIn$ in Figure 1) is defined as the average number of bibliographic coupling links per cited reference of the FP's citing papers. In contrast to the other indicators mentioned earlier, $DeIn$ is supposed to decrease with the disruptiveness of a paper, since it measures the dependency of the paper on earlier work (as opposed to disruptiveness). Another difference to the other indicators is that the range of $DeIn$ has no upper bound, since the average number of bibliographic coupling links between the FP and its citing papers is not limited. This makes it more difficult to compare the results of $DeIn$ and the other indicators.

By considering those cited references of the FP's citing papers with at least $l$ ($l > 1$) bibliographic coupling links with the FP, it becomes possible to follow the idea of taking into account how strong the cited references of the FP's citing papers rely on the cited references of the FP, but also get values which are more comparable to the other indicators. This is formalized in the formulae in Figure 1, where the subscripts of $DI_l$ and $DI_1^{no\ k}$ correspond to the threshold for the number of bibliographic coupling links. With a threshold of $l = 1$ (i.e., without any restriction on the number of bibliographic coupling links), the indicator is identical to the indicator originally proposed by Wu et al. (2019). In order to analyze how well these different strategies are able to measure the disruptiveness of a paper, we compare the following indicators in our analyses: $DI_1$, $DI_5$, $DI_1^{no\ k}$, $DI_5^{no\ k}$, $DeIn$. The subscript in four variants indicates the minimum number of cited references that are cited along with the focal paper. The superscript "no k" in two variants indicates that $N_k$ is excluded from the calculation.

# 3 Methods

## 3.1 F1000Prime

F1000Prime is a database including important papers from biological and medical research (see https://f1000.com/prime/home). The database is based on a post-publication peer review system: peer-nominated Faculty Members (FMs) select the best papers in their specialties and assess these papers for inclusion in the F1000Prime database. FMs write brief reviews explaining the importance of papers and rate them as "good" (1 star), "very good" (2 stars) or "exceptional" (3 stars). Many papers in the database are assessed by more than one FM. To rank the papers in the F1000Prime database, the individual scores are summed up to a total score for each paper.

FMs also assign the following tags to the papers, if appropriate:[1]

- Confirmation: article validates previously published data or hypotheses
- *Controversial*: article challenges established dogma
- Good for teaching: key article in a field and/or is well written
- **Hypothesis**: article presents an interesting hypothesis
- Negative/null results: article has null or negative findings
- **New finding**: article presents original data, models or hypotheses
- **Novel drug target**: article suggests new targets for drug discovery
- Refutation: article disproves previously published data or hypotheses
- **Technical advance**: article introduces a new practical/theoretical technique, or novel use of an existing technique

The tags in bold reflect aspects of novelty in research. Since disruptive research should include elements of novelty, we expect that the tags are positively related to the disruption indicator scores. For instance, we assume that a paper receiving many "new finding" tags from FMs will have a higher disruption index score than a paper receiving only a few tags (or none at all). The tags not printed in bold are not related to newness (e.g., confirmation of previously published hypotheses), so that the expectations for these tags are zero or negative correlations with disruption index scores. The tag "controversial" is printed in italics. It is not clear whether the tag is able to reflect novelty or not. FMs further assign the tags "clinical trial", "systematic review/meta-analysis", and "review/commentary" to papers which are not relevant for this study (and not used thus).

---

[1] The definitions of the tags are adopted from https://f1000.com/prime/about/whatis/how

Many other studies have already used data from the F1000Prime database for correlating them with metrics. Most of these studies are interested in the relationship between quantitative (metrics-based) and qualitative (human-based) assessments of research. The analysis of Anon (2005) shows that "papers from high-profile journals tended to be rated more highly by the faculty; there was a tight correlation ($R^2 = 0.93$) between average score and the 2003 impact factor of the journal" (see also Jennings, 2006). Bornmann and Leydesdorff (2013) correlated several bibliometric indicators and F1000Prime recommendations. They found that the "percentile in subject area achieves the highest correlation with F1000 ratings" (p. 286). Waltman and Costas (2014) report "a clear correlation between F1000 recommendations and citations. However, the correlation is relatively weak" (p. 433). Similar results were published by Mohammadi and Thelwall (2013). Bornmann (2015) investigated the convergent validity of F1000Prime assessments. He found that "the proportion of highly cited papers among those selected by the faculty members is significantly higher than expected. In addition, better recommendation scores are also associated with higher performing papers" (p. 2415). The most recent study by Du, Tang, and Wu (2016) show that "(a) nonprimary research or evidence-based research are more highly cited but not highly recommended, while (b) translational research or transformative research are more highly recommended but have fewer citations" (p. 3008).

**3.2 Dataset used and variables**

The study is based on a dataset from F1000Prime including 207,542 assessments of papers. These assessments refer to 157,020 papers (excluding papers with duplicate DOIs, missing DOIs, missing expert assessments etc.). The bibliometric data for these papers are from an in-house database (Korobskiy, Davey, Liu, Devarakonda, & Chacko, 2019), which utilizes Scopus data (Elsevier Inc.). To increase the validity of the indicators included in this study, we considered only papers with at least 10 cited references and at least 10 citations. Furthermore, we included only papers from 2000 to 2016 to have reliable data (some publications are from 1970 or earlier) and a citation window for the papers of at least three years (since publication until the end of 2018). The reduced paper set consists of 120,179 papers published between 2000 and 2016 (see Table 1).

Table 1. Number and percentage of papers included in the study

| Publication year | Number of papers | Percentage of papers |
|---|---|---|
| 2000 | 196 | 0.16 |
| 2001 | 1,530 | 1.27 |
| 2002 | 3,229 | 2.69 |
| 2003 | 3,717 | 3.09 |
| 2004 | 5,185 | 4.31 |
| 2005 | 6,711 | 5.58 |
| 2006 | 8,765 | 7.29 |
| 2007 | 8,824 | 7.34 |
| 2008 | 10,046 | 8.36 |
| 2009 | 10,368 | 8.63 |
| 2010 | 11,074 | 9.21 |
| 2011 | 10,934 | 9.1 |
| 2012 | 10,536 | 8.77 |
| 2013 | 9,903 | 8.24 |
| 2014 | 7,261 | 6.04 |
| 2015 | 6,121 | 5.09 |
| 2016 | 5,779 | 4.81 |
| Total | 120,179 | 100.00 |

We included several variables in the empirical part of this study: the disruption index proposed by Wu et al. (2019) ($DI_1$) and the dependence indicator proposed by Bu et al. (2019)

($DeIn$). The alternative disruption indicators described in section 2 considered were: $DI_5$, $DI_1^{no\ k}$, and $DI_5^{no\ k}$. For the comparison with the indicators reflecting disruption, we included the sum (ReSc.sum) and the average (ReSc.avg) of reviewer scores (i.e., scores from FMs). Besides the qualitative assessments of research, quantitative citation impact scores are also considered: number of citations until the end of 2018 (Citations) and percentile impact scores (Percentiles).

Since publication and citation cultures are different in the fields, it is standard in bibliometrics to field- and time-normalize citation counts (Hicks, Wouters, Waltman, de Rijcke, & Rafols, 2015). Percentiles are field- and time-normalized citation impact scores (Bornmann, Leydesdorff, & Mutz, 2013) that are between 0 and 100 (higher scores reflect more citation impact). For the calculation of percentiles, the papers published in a certain subject category and publication year are ranked in decreasing order. Then the formula $(i - 0.5)/n * 100$ (Hazen, 1914) is used to calculate percentiles ($i$ is the rank of a paper and $n$ the number of papers in the subject category). Impact percentiles of papers published in different fields can be directly compared (despite possibly differing publication and citation cultures).

Table 2. Key figures of the included variables (*n*= 120,179)

| Variable | Mean | Median | Standard deviation | Minimum | Maximum |
|---|---|---|---|---|---|
| $DI_1$ | -0.007 | -0.004 | 0.013 | -0.322 | 0.677 |
| $DI_5$ | 0.089 | -0.007 | 0.278 | -0.800 | 1.000 |
| $DI_1^{no\ k}$ | -0.521 | -0.579 | 0.294 | -0.998 | 0.975 |
| $DI_5^{no\ k}$ | -0.008 | -0.053 | 0.545 | -0.990 | 1.000 |
| $DeIn$ | 3.327 | 2.970 | 1.871 | 0.013 | 43.059 |
| ReSc.sum | 2.028 | 2.000 | 1.808 | 1.000 | 55.000 |
| ReSc.avg | 1.486 | 1.000 | 0.586 | 1.000 | 3.000 |
| Citations | 149.848 | 73.000 | 298.467 | 10.000 | 20446.000 |
| Percentiles | 87.246 | 91.947 | 13.248 | 23.659 | 100.000 |

Table 2 shows the key figures for citation impact scores, reviewer scores, and variants measuring disruption. As the percentiles reveal, the paper set includes especially papers with a considerable citation impact. The maximum $DI_1$ with the value 0.677 has been reached by the paper entitled "Cancer statistics, 2010" published by Jemal, Siegel, Xu, and Ward (2010). This publication is one of an annual series published on incidence, mortality, and survival rates for cancer and its high score may be an artifact of the $DI_1$ formula since it is likely that the report is cited much more than its cited references. In fact, this publication may make the case for the $DI_5$ formulation.

### 3.3   Statistics applied

The statistical analyses in this study have three steps:

(1) We investigated the correlations between citation impact scores, reviewer scores, and the scores of the indicators measuring disruption. All variables are not normally distributed and affected by outliers. In order to tackle this problem, we logarithmized the scores by using the formula log($x$+1). This logarithmic transformation approximates the distributions to normal distributions. Since perfectly normally distributed variables cannot be achieved with the transformation, Spearman rank correlations have been calculated (instead of Pearson correlations). We interpret the correlation coefficients against the backdrop of the guidelines proposed by Cohen (1988) and Kraemer et al. (2003): small effect=0.1, medium effect=0.3, large effect=0.5, and very large effect=0.7.

(2) We performed an exploratory factor analysis (FA) to analyze the variables. FA is a statistical method for data reduction (Gaskin & Happell, 2014); it is an exploratory technique to identify latent dimensions in the data and to investigate how the variables are related to the dimensions (Baldwin, 2019). We expected three dimensions, because we have variables with citation impact scores, reviewer scores, and indicators' scores measuring disruption. Since the (logarithmized) variables do not perfectly follow the normal distribution, we performed the FA using the robust covariance matrix following Verardi and McCathie (2012). Thus, the results of the FA are not based on the variables, but on a covariance matrix. The robust covariance matrix has been transformed into a correlation matrix (StataCorp., 2017), which has been analyzed by the principal-component factor method (the communalities are assumed to be 1). We interpreted the factor loadings for the orthogonal varimax rotation; the factor loadings have been adjusted "by dividing each of them by the communality of the correspondence variable. This adjustment is known as the Kaiser normalization" (Afifi, May, & Clark, 2012, p. 392). In the interpretation of the results, we focused on factor loadings with values greater than 0.5.

(3) We investigated the relationship between the dimensions (identified in the FA) and F1000Prime tags (as proxies for newness or not). We expected a close relationship between the dimension reflecting disruption and tags reflecting newness. The tags are count variables including the sum of the tags assignments from F1000Prime FMs for single papers. For the calculation of the relationship between dimensions and tags, we performed a robust Poisson regression (Hilbe, 2014; Long & Freese, 2014). The Poisson model is recommended to be used in cases of count data as dependent variable. Robust methods are recommended when the distributional assumptions for the model are not completely met (Hilbe, 2014). Since we are interested in identifying indicators for measuring disruption which might perform better than the other variants, we tested the correlation between each variant and the tag assignments using several robust Poisson regressions. Citations, disruptiveness, and tag assignments are dependent on time (Bornmann & Tekles, 2019). Thus, we included the number of years between 2018 and the publication year as exposure time in the models (Long & Freese, 2014, pp. 504-506).

# 4 Results

## 4.1 Correlations between citation impact scores, reviewer scores, and variants measuring disruption

Figure 2 shows the matrix including the coefficients of the correlations between reviewer scores, citation impact indicators, and variants measuring disruption. $DI_1$ is correlated on a medium level with the other indicators measuring disruption whereby these indicators correlate among themselves on a very high level. Very high positive correlations are visible between citations and percentiles and between the average and sum of reviewer scores.

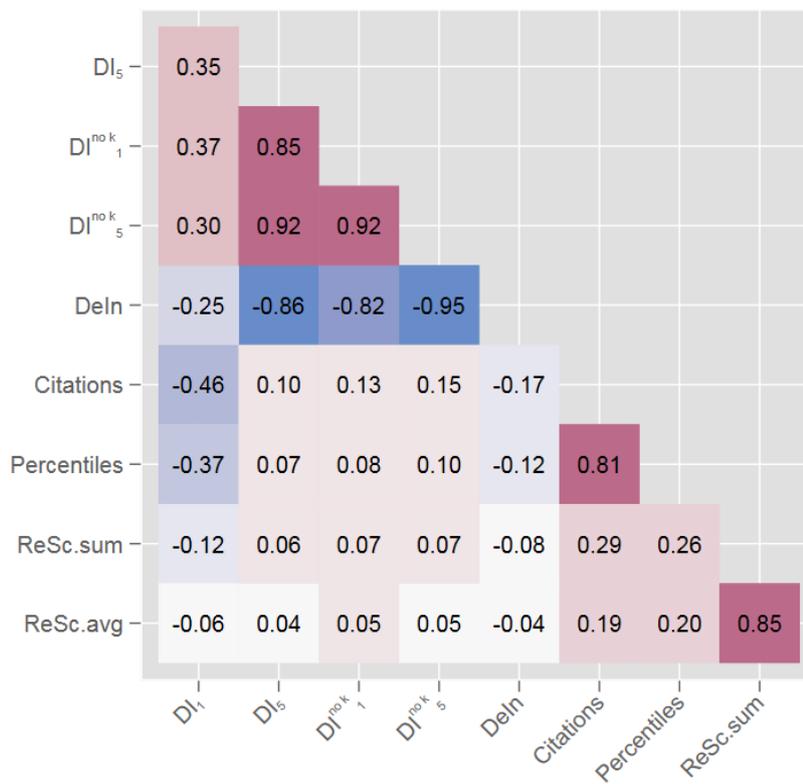

Figure 2. Spearman rank correlations based on logarithmized variables [log(y+1)]. The following abbreviations are used: different indicators measuring disruption ($DI_1$, $DI_5$, $DI_1^{no\ k}$, $DI_5^{no\ k}$, $DeIn$), the sum (ReSc.sum) and the average (ReSc.avg) of reviewer scores.

The correlation between $DI_1$ and citation impact (citations and percentiles) is at least on the medium level but it is negative (r=-0.46, r=-0.37). Thus, the original $DI_1$ seems to measure another dimension than citation impact. This result is in agreement with results reported by Wu et al. (2019, Figure 2a). However, the situation changed with the other indicators measuring disruption to small positive (negative in the case of $DeIn$) correlation coefficients.

### 4.2 Factor analysis to identify latent dimensions

We calculated a factor analysis including reviewer scores, citation impact indicators, and variants measuring disruption to investigate the underlying dimensions (latent variables). Most of the results that are shown in Table 3 agree with the expectations: we found three dimensions which we labeled as disruption (factor 1), citations (factor 2), and reviewers (factor 3). However, other than expected, $DI_1$ loads negatively on the citation dimension revealing that (1) high $DI_1$ scores are related to low citation impact scores (see above) and (2) all other indicators measuring disruption are independent of $DI_1$. Thus, the other indicators (at least one) seem to be promising developments compared to the originally proposed indicator $DI_1$.

Table 3. Rotated factor loadings from a factor analysis using logarithmized variables [log(y+1)]

| Variable | Factor 1 | Factor 2 | Factor 3 | Uniqueness |
|---|---|---|---|---|
| $DI_1$ | 0.24 | **-0.69** | 0.05 | 0.46 |
| $DI_5$ | **0.90** | -0.07 | 0.00 | 0.19 |
| $DI_1^{no\ k}$ | **0.90** | -0.10 | 0.02 | 0.17 |
| $DI_5^{no\ k}$ | **0.97** | -0.03 | 0.01 | 0.05 |
| $DeIn$ | **-0.91** | -0.01 | 0.01 | 0.17 |
| Citations | 0.05 | **0.91** | 0.04 | 0.16 |
| Percentiles | 0.04 | **0.84** | 0.12 | 0.29 |
| ReSc.sum | 0.00 | 0.05 | **1.00** | 0.00 |
| ReSc.avg | 0.00 | 0.05 | **1.00** | 0.00 |

Notes. Three eigenvalues > 1. The following abbreviations are used: different indicators measuring disruption ($DI_1$, $DI_5$, $DI_1^{no\ k}$, $DI_5^{no\ k}$, and $DeIn$), the sum (ReSc.sum) and the average (ReSc.avg) of reviewer scores.

### 4.3 Relationship between tag mentions and FA dimensions

Using Poisson regression models including the tags, we calculated correlations between the tags and the three FA dimensions (disruption, citations, and reviewers). We are especially interested in the correlation between the tags (measuring newness of research or not) and the disruption dimension from the FA. We also included the citation impact and reviewers dimensions into the analyses to see the corresponding results for comparison. In the analyses, we considered the FA scores for the three dimensions (predicted values from FA which are not correlated by definition) as independent variables and the various tags as dependent variables in nine Poisson regressions (one regression model for each tag).

Table 4. Results of nine Poisson regression analyses (*n*=120,179 papers). The models have been adjusted for exposure time (different publication years): how long was the time that the papers have been at risk of being tagged and cited (number of years between publication and counting of citations or tags, respectively)?

| Tag | Disruption | | Citations | | Reviewers | | Constant |
|---|---|---|---|---|---|---|---|
| | Coefficient | Percentage change | Coefficient | Percentage change | Coefficient | Percentage change | |
| Tags expressing newness (expecting positive signs) | | | | | | | |
| Hypothesis | -0.06*** | -6.11 | 0.20*** | 28.62 | 0.32*** | 41.28 | -3.96*** |
| | (-10.22) | | (36.91) | | (33.24) | | (-36.32) |
| New finding | -0.09*** | -9.92 | 0.23*** | 34.02 | 0.27*** | 33.48 | -3.35*** |
| | (-41.77) | | (111.06) | | (76.41) | | (-83.59) |
| Novel drug target | -0.01 | -1.66 | 0.27*** | 39.91 | 0.34*** | 43.93 | -5.87*** |
| | (-1.45) | | (27.33) | | (22.96) | | (-34.41) |
| Technical advance | 0.09*** | 10.93 | 0.22*** | 32.22 | 0.25*** | 30.74 | -4.58*** |
| | (13.67) | | (32.56) | | (24.34) | | (-39.50) |
| Tags not expressing newness (expecting negative signs) | | | | | | | |
| Confirmation | -0.03*** | -3.85 | 0.14*** | 19.69 | 0.13*** | 14.30 | -4.56*** |
| | (-6.41) | | (28.93) | | (21.86) | | (-68.84) |
| Good for teaching | -0.06*** | -7.08 | 0.14*** | 19.00 | 0.38*** | 49.61 | -4.06*** |
| | (-6.73) | | (14.71) | | (19.75) | | (-21.12) |
| Negative/Null results | -0.14** | -14.72 | 0.17*** | 23.23 | 0.34*** | 44.25 | -7.86*** |
| | (-3.25) | | (5.09) | | (10.02) | | (-19.94) |
| Refutation | -0.19*** | -19.00 | 0.23*** | 32.88 | 0.28*** | 34.47 | -7.71*** |
| | (-8.14) | | (12.05) | | (12.64) | | (-28.50) |
| Tag without expectations | | | | | | | |
| Controversial | -0.04*** | -4.78 | 0.20*** | 28.11 | 0.27*** | 33.54 | -5.30*** |
| | (-4.54) | | (24.85) | | (27.92) | | (-47.46) |

Notes. *t* statistics in parentheses; * *p* < 0.05, ** *p* < 0.01, *** *p* < 0.001. Percentage change coefficients in green are as expected; red coefficients flag results against expectations.

The results of the regression analyses are shown in Table 4. We do not focus on the statistical significance of the results, since they are more or less meaningless against the backdrop of the high case numbers. The most important information in the table are the signs of the coefficients and the percentage changes coefficients. The percentage change coefficients are counterparts to odds ratios in regression models which measure the percent changes in the dependent variable if the independent variable (FA score) increases by one standard deviation (Deschacht & Engels, 2014; Hilbe, 2014; Long & Freese, 2014). The percentage change coefficient for the model based on the "technical advance" tag and the disruption dimension can be interpreted as follows: for a standard deviation increase in the scores for disruption, a paper's expected number of new finding tags increases by 10.93%, holding other variables in the regression analysis constant. This increase is as expected and substantial. However, the results of the other tags expressing newness have a negative sign and are against the expectation.

The percentage change coefficients for the citations dimensions are significantly higher than that for the disruption dimension (especially for the new finding tag) and positive. This result might be against our expectations, since the disruption variants should measure newness in a better way than citations. However, one should consider in the interpretation of the results that $DI_1$ correlates negatively with the citation indicators. Thus, the dimension also measures disruptiveness (as originally proposed) whatever the case may be. If we interpret the results for the dimension against this backdrop, they seem to accord with the expectation for disruptiveness. However, for the tags not expressing newness, the percentage change coefficients are also positive, which is against the expectation. The results for the reviewers dimension are similar to the citations dimension results. The consistent positive coefficients for the citations and reviewers dimensions in Table 4 might result from the fact that the tags are from the same FMs as the recommendations, and the FMs probably use citations to find relevant papers for reading, assessing, and including in the F1000Prime database.

Table 5. Results (percentage change coefficients) of 45 Poisson regressions with tags as dependent variables and different variants measuring disruption as independent variables each. The models have been adjusted for exposure time (different publication years): how long have the papers been at risk of being tagged and cited?

| Tag | $DI_1$ | $DI_5$ | $DI_1^{no\ k}$ | $DI_5^{no\ k}$ | $DeIn$ |
|---|---|---|---|---|---|
| Tags expressing newness (expecting positive signs) | | | | | |
| Hypothesis | 4.32 | 3.01 | 4.66 | 2.75 | 0.25 |
| New finding | -2.71 | -0.62 | -2.13 | -2.13 | 1.92 |
| Novel drug target | 6.89 | 6.74 | 14.85 | 15.19 | -7.91 |
| Technical advance | 6.72 | 20.65 | 18.34 | 19.80 | -18.11 |
| Tags not expressing newness (expecting negative signs) | | | | | |
| Confirmation | -5.08 | -0.41 | 0.08 | 1.45 | -1.88 |
| Good for teaching | 12.24 | 0.37 | 8.82 | 3.73 | 6.41 |
| Negative/Null results | 3.45 | -11.46 | -13.35 | -5.20 | -2.10 |
| Refutation | -9.28 | -9.59 | -14.83 | -10.37 | 8.06 |
| Tag without expectations | | | | | |
| Controversial | -0.33 | 4.42 | 1.46 | 1.46 | -3.62 |
| | | | | | |
| Results meeting the expectations | 5 | 6 | 5 | 5 | 4 |

Note. The following abbreviations are used: different indicators measuring disruption ($DI_1$, $DI_5$, $DI_1^{no\ k}$, $DI_5^{no\ k}$, $DeIn$). Percentage change coefficients in green are as expected; red coefficients flag results against expectations.

Table 5 reports the results from some additional regression analyses. Since we are not only interested in correlations between dimensions (reflecting disruptiveness) and tags, but also in correlations between the various variants measuring disruption and tags, we calculated 45 additional regression models. We are interested in the question which variant measuring disruption reflect newness better than other variants: are the different variants differently or similarly related to newness – as expressed by the tags? Table 5 only shows percentage change coefficients (see above) from the regression models (because of the great number of models). In other words, percentage changes in expected counts (of the tag) for a standard deviation increase in the variant measuring disruption are listed. For example, a standard deviation change in $DI_1$ on average increases a paper's expected number of technical advance tags by 6.72%. This result agrees to the expectation, since the technical advance tag reflects newness.

The last line in Table 5 shows the number of percent change coefficients for an indicator being as expected. It seems that $DI_5$ reflects the assessments by FMs at best; the lowest number of results in agreement is visible for $DeIn$.

## 5    Discussion

For many years, scientometrics research has focused on improving the way of field-normalizing citation counts or developing improved variants of the h index. However, this research is rooted in relatively one-dimensional way of measuring impact. With the introduction of the new family of disruption indicators, the one-dimensional way of impact measurement may now give way to multi-dimensional approaches. Disruption indicators consider not only times-cited information, but also cited references data (of FPs and citing papers). High indicator values should be able to point to published research disrupting traditional research lines. Disruptive papers catch the attention of citing authors (at the expense of the attention devoting to previous research); disruptive research enters unknown territory which is scarcely consistent with known territories handled in previous papers (and cited by disruptive papers). Thus, the citing authors exclusively focus on the disruptive papers (by citing them) and do not reference previous papers cited in the disruptive papers.

Starting from the basic approach of comparing cited references of citing papers with cited references of FPs, different variants of measuring disruptiveness have been proposed recently. An overview of many possible variants can be found in Wu and Yan (2019). In this study, we included some variants which sounded reasonable and/or followed different approaches. For example, $DeIn$ proposed by Bu et al. (2019) is based on bibliographic coupling links (without considering $N_k$). We were interested in the convergent validity of these new indicators: following the basic analyses approach by Bornmann et al. (in press), we wanted to know whether these indicators measuring disruptiveness are really able to measure what they propose to measure. The convergent validity can be tested by using an external criterion measuring the same dimension. Although we did not have an external criterion at hand measuring disruptiveness specifically, we used tags from the F1000Prime database reflecting newness. FMs assess papers using certain tags and some tags reflect newness. We assumed that disruptive research is assessed as new. Based on the F1000Prime data, we investigated whether the tags assigned by the FMs to the papers correspond with indicator values measuring disruptiveness.

In the first step of the statistical analyses, we calculated a FA for inspecting whether the various indicators measuring disruptiveness load on a single 'disruptiveness' dimension. As the results reveal, this was partly the case: all variants of the $DI_1$ – the original disruption index proposed by Wu et al. (2019) – loaded on one dimension – the 'disruptiveness' dimension. However, the original disruption index itself loaded on a dimension which reflects citation impact; however, it loaded negatively. These results might be interpreted as follows: the proposed disruption index variants measure the same construct which might be interpreted as 'disruptiveness'. $DI_1$ is related to citation impact whereby negative values – the developmental index manifestation of this indicator (see section 2) – correspond to high citation impact levels. Since all variants of $DI_1$ loaded on the same factor in the FA, the results do not show which variant should be preferred (if any). Thus, we considered a second step of analyses in this study.

In this step, we tested the correlation between each variant (including the original) and the external 'newness' criterion. The results showed that $DI_5$ reflects the FMs' assessments at best (corresponds with our expectations more frequently than the other indicators); the lowest number of results which demonstrated an agreement between tag and indicator scores is visible for $DeIn$. The difference between the variants is not very large; however, the results can be used to guide the choice of a variant if disruptiveness is measured in a scientometric study. Although the authors of the paper introducing $DI_1$ (Wu et al., 2019) performed analyses to validate the index (e.g., by calculating the indicator for Nobel-prize-winning papers), they did not report on evaluating possible variants of the original which might perform better.

We noted that while a single publication was the most highly disruptive for the $DI_1$ (0.6774), and $DI_1^{no\ k}$ (0.9747), 703 and 3816 publications respectively scored at the maximum disruptiveness value of 1.0 for variants $DI_5$ and $DI_5^{no\ k}$. We also reviewed examples of the most highly disruptive publications as measured by all four variants and observed that instances of an annual Cancer Statistics report published by the American Cancer Society received maximal disruptiveness scores for all four variants presumably because this reports is highly cited in each year of its publication without its references being cited. A publication from the *Journal of Global Environmental Change* (10.1016/j.gloenvcha.2008.10.009) was also noteworthy and may reflect focus on climate change.

It would be interesting to follow up in future studies that use mixed-methods approaches to more systematically evaluate the properties of $N_i$, $N_j$, and $N_k$ variants against additional gold standard datasets. The F1000 dataset is certain to feature its own bias (e.g., it is restricted to biomedicine and includes disproportionally many high-impact papers) and the variants we describe may exhibit different properties when evaluated against multiple datasets.

# Acknowledgements

We would like to thank Tom Des Forges and Ros Dignon from F1000 for providing us with the F1000Prime dataset. Research and development reported in this publication was partially supported by funds from the National Institute on Drug Abuse, National Institutes of Health, US Department of Health and Human Services, under Contract No HHSN271201800040C (N44DA-18-1216).

# Competing Interests

The authors do not declare any competing interests. Citation data used in this paper relied on Scopus (Elsevier Inc.) as implemented in the ERNIE project (Korobskiy et al., 2019), which is collaborative between NET ESolutions Corporation and Elsevier Inc. The content of this publication is solely the responsibility of the authors and does not necessarily represent the official views of the National Institutes of Health, NET ESolutions Corporation, or Elsevier Inc.